\documentstyle[11pt,paspconf,epsf]{article}

\def\gtorder{\mathrel{\raise.3ex\hbox{$>$}\mkern-14mu
             \lower0.6ex\hbox{$\sim$}}}
\def\ltorder{\mathrel{\raise.3ex\hbox{$<$}\mkern-14mu
             \lower0.6ex\hbox{$\sim$}}}

\begin{document}
\title{ The Interstellar Medium of Lens Galaxies}
\author{B. A. McLeod, E. E. Falco, C. S. Kochanek, J. Leh\'ar,
J. A. Mu\~noz} 
\affil{Harvard-Smithsonian Center for Astrophysics\\ 
           60 Garden St., Cambridge, MA 02138}
\author{C. D. Impey, C. Keeton, C. Y. Peng} 
\affil{Steward Observatory, University of Arizona\\
           Tucson, AZ 85721}
\author{H.-W. Rix}
\affil{Max-Planck Institut f\"ur Astrophysik, Heidelberg, Germany}
 
\begin{abstract}
We use observations from the CASTLES survey of gravitational lenses
to study extinction in 23 lens galaxies with $0 < z_l < 1$. The median
differential extinction between lensed images is $\Delta E(\bv) = 0.05$
mag, and the directly measured extinctions agree with the amount needed to
explain the differences between the statistics of radio and (optical)
quasar lens surveys.  We also measure the first extinction laws outside
the local universe, including an $R_V=7.2$ curve for a molecular cloud
at $z_l=0.68$ and an $R_V=1.5$ curve for the dust in a redshift $z_l=0.96$
elliptical galaxy.
\end{abstract}


\section{Introduction}

Extinction corrections are important for determining the Hubble
Constant (e.g. Freedman et al. 1998), the cosmological model
(e.g. Perlmutter et al. 1997, Riess et al. 1998, Falco, Kochanek, \&
Mu\~noz 1998), and the epoch of star formation (e.g. Madau, Pozetti,
\& Dickinson 1998), but detailed studies of extinction are limited to
nearby galaxies where individual stars can be observed.  At larger
distances, one must rely on studies of the spectral energy
distribution of stars mixed with dust in an uncertain geometry.
Gravitational lensing eliminates this uncertainty by using multiple
lines of sight through a galaxy to the same background source.  
Thus, lensing is a powerful tool to measure the extinction properties of
galaxies at moderate redshift.

The variation of absorption, $A_\lambda$, as a function of wavelength
is described by the extinction law $R_\lambda$ where
$A_\lambda=R_\lambda E(\bv)$.  Galactic extinction laws are well
modeled by parametrized functions of $R_V$, where we have used the
Cardelli et al. (1989) models.  The typical Galactic value is
$R_V=3.1$, but the overall range is $2.2 \ltorder R_V \ltorder 5.8$.
The value of $R_V$ depends on the size and composition of the dust
along the line of sight, and in the SMC and some regions of the LMC
the UV extinction curves show large deviations from the Galactic
models. The properties of local dust are reviewed by Mathis (1990).

We consider here the multiply-imaged quasars observed for the Center
for Astrophysics/Arizona Space Telescope Lens Survey (CASTLES, see Falco
et al. 1999b), combined with archival HST data.   Our initial
survey of extinction in 23 lenses, 8 of which were radio selected,
is discussed by Falco et al. (1999a).  

\section{Method}

We measure the differential extinction between lines of sight in a
gravitational lens by measuring the difference in color between the
several images in the lens.  In our analysis we make a number of
simplifying assumptions: (1) the magnification is wavelength
independent; (2) the magnification is time independent; (3) the source
spectrum is time independent and (4) the extinction law is identical
along the various lines of sight.  Then the magnitude difference
between images $i$ and $j$ is
\begin{equation}
 m_i(\lambda)-m_j(\lambda) = - 2.5 \log { M_i \over M_j} + \Delta E_{ij}
 R \left( { \lambda \over 1+z_l }\right),
\end{equation}
which depends only on the magnification ratio, $M_i/M_j$, the extinction difference 
between the two lines of sight, $\Delta E_{ij}$, and the extinction law $R(\lambda/(1+z_l))$ 
in the lens rest frame.  This dependence of the magnitude difference on 
extinction was first explored by Nadeau et al. (1991).

\section{Results}

\begin{figure}[p]
\plotone{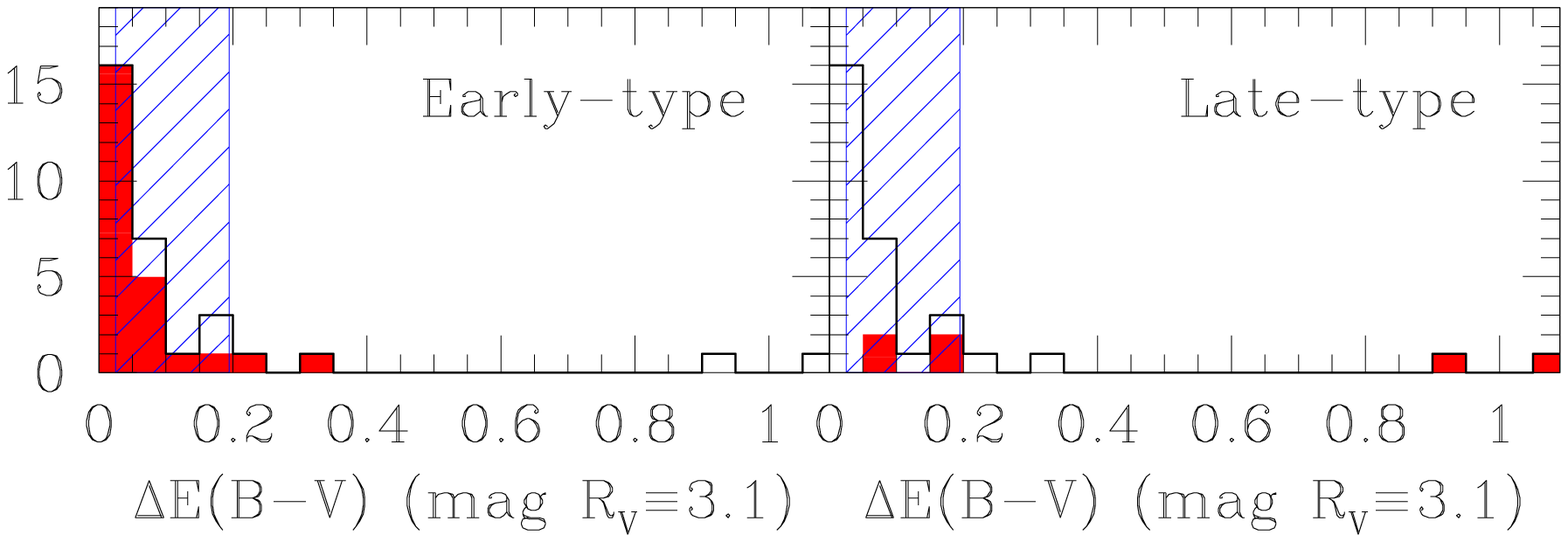}
\caption{\label{fig-differential} Histograms of the differential
extinction in early type (left, shaded) and late type (right, shaded) lens 
galaxies.  The superposed unshaded histogram shows the total distribution.
The hatched region shows the extinction range estimated
from a comparison of the statistics of optically-selected and radio-selected
lenses by Falco et al. (1998).  }

\bigskip

\plottwo{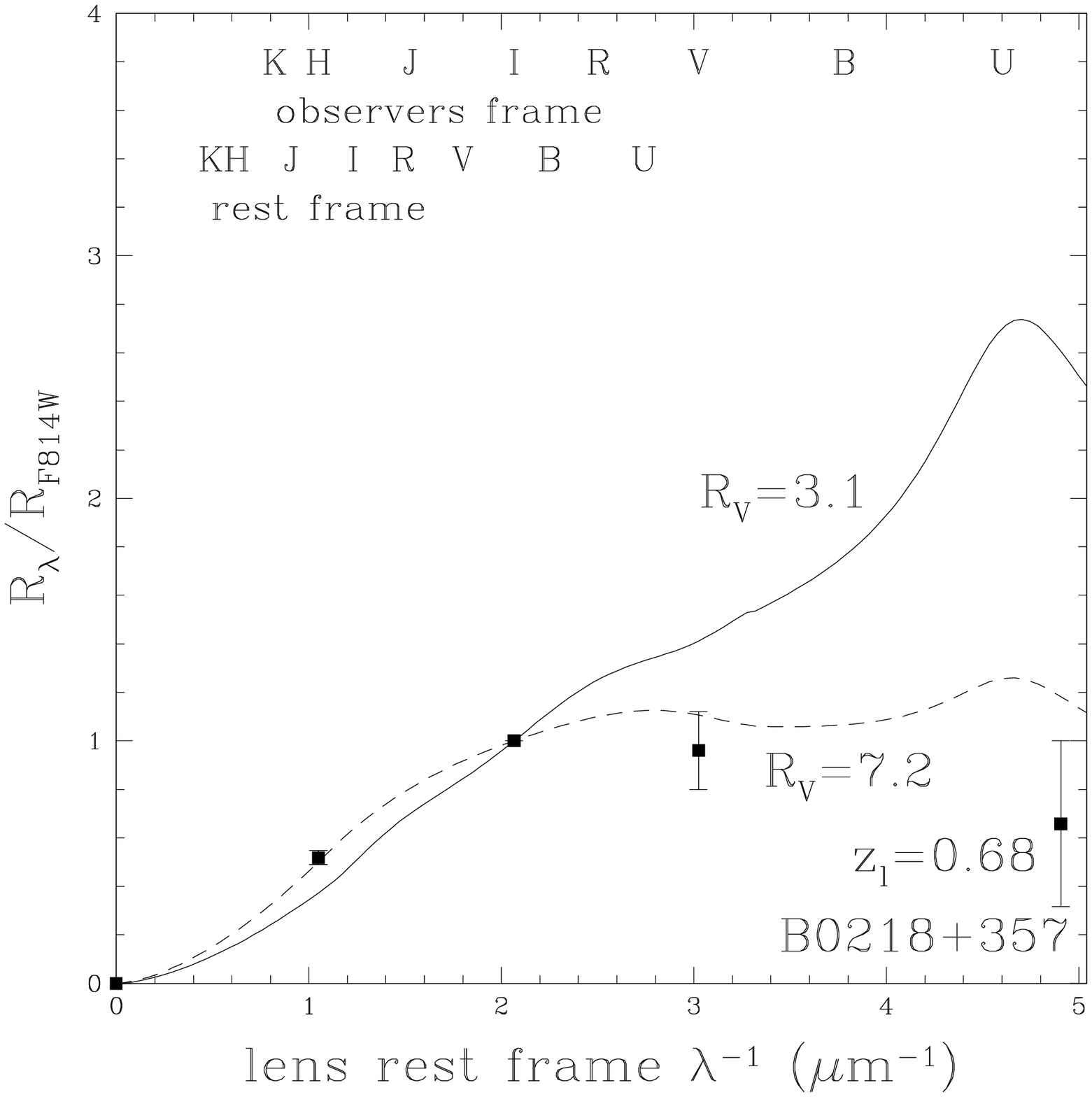}{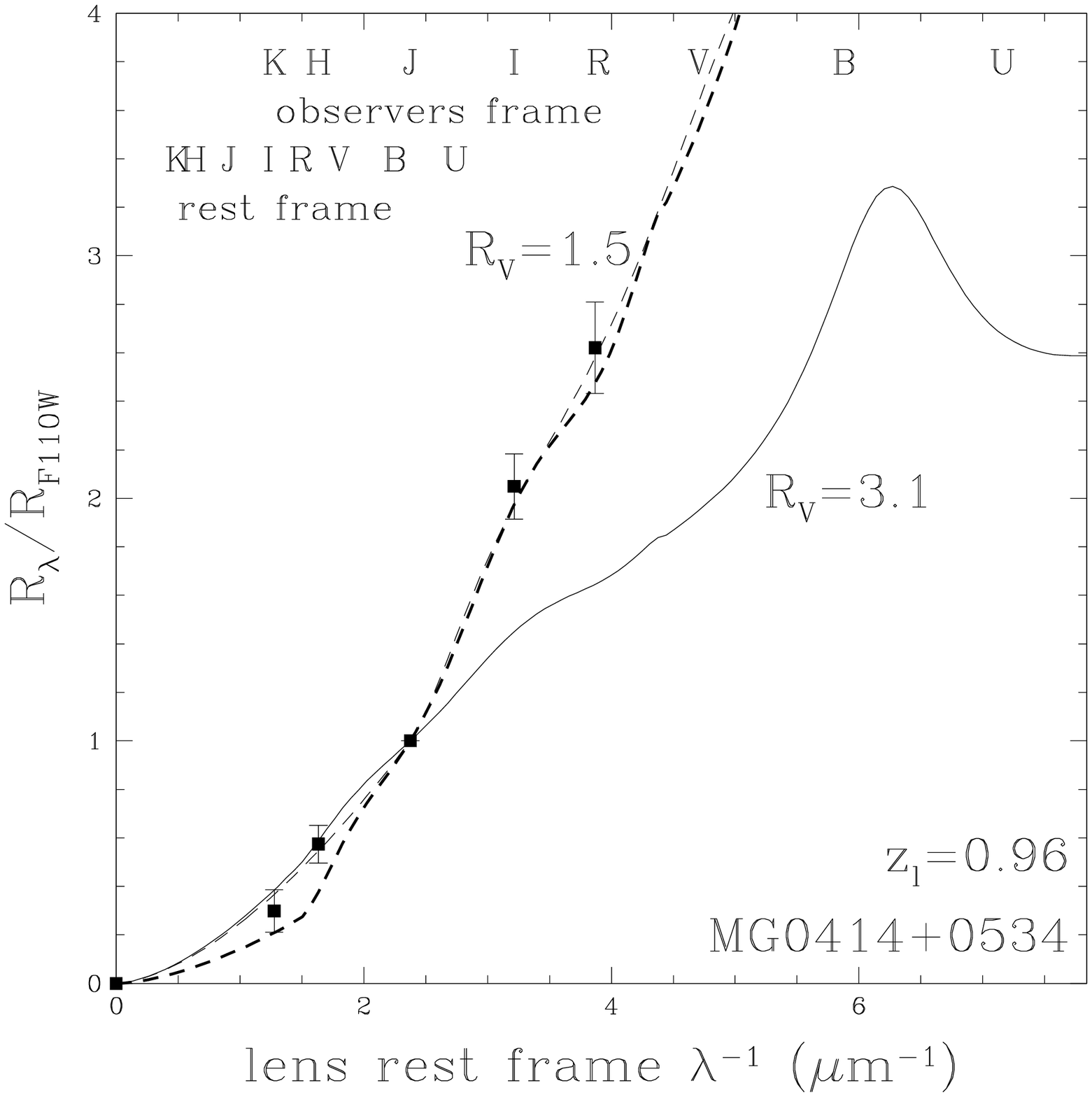}

\caption{ \label{fig-curve} Extinction curves for B~0218+357 and
MG~0414+0534.  The solid line shows the standard $R_V=3.1$ curve and
the dashed line shows the best fit parametric curve.  For simplicity
the curves are normalized by the $R_\lambda$ value of the filter
closest to the lens rest frame V band.  Both lensed sources are radio
emitters and include a point at $\lambda^{-1}=0$, allowing an absolute
determination of both the extinction and the extinction curve. }
\end{figure}

We initially determine values for $\Delta E_{ij}$ by assuming a
standard Galactic $R_V=3.1$ extinction curve and then fitting the
image magnitude differences as a function of wavelength for each
lens. Figure~\ref{fig-differential} shows the distribution of
differential extinctions for both early and late type galaxies.  Most
lens galaxies are early type galaxies, but the two lenses with the
highest differential extinction, PKS~1830-211 and B~0218+357, are both
late type galaxies where one of the lensed images lies behind a
molecular cloud (Menten \& Reid 1996; Frye et al. 1997).  In both
cases, we find dust-to-gas ratios a factor of 2--5 below standard
estimates.  The median differential extinction is $\Delta E(\bv)=0.04$
mag for optically selected lenses and $0.06$ mag for radio-selected
lenses.  This difference is not surprising, as we expect a bias against
dusty lenses in optically selected samples.

We can determine the total extinctions only by using a model for the
intrinsic spectra of quasars.  While we can determine differential
extinctions with an accuracy of $\sim 0.02$ mag, we can measure total
extinctions only with an accuracy of $\sim 0.1$ mag.  Formally we find
that the median total extinction of the bluest images is $E(\bv)$ =
0.08\,mag.  This matches the mean extinction of $E(\bv)=0.10\pm0.08$ mag
required to explain the differences between the statistics of radio
and optically selected lens samples (Falco et al. 1998).  Extinctions
at the level of $E(\bv)\sim0.1$ mag lead to significant corrections to the
statistics of optically selected lenses, and must be included when using
optically selected samples to determine the cosmological model.

Of the 23 systems considered here, 12 have adequate data to 
estimate both $\Delta E$ and $R_V$.  Of these, 7 are consistent
with the Galactic value of $R_V$=3.1, and the remaining 5 
have values ranging from 1.5 and 7.2.  Figure~\ref{fig-curve}
shows the extinction curve derived for the two extreme cases. In  
B0218+357, a $z=0.68$ late-type galaxy where one image passes
through a molecular cloud,  we find a very high value of $R_V=7.2$.  
This is similar to the Galaxy where high values of $R_V$ are associated with
denser regions of the ISM.  In MG0414+534, a $z=0.96$ elliptical, we
find $R_V=1.5$.  

While we see dust in lenses, our overall results rule out the ``dusty
lens'' hypothesis of Malhotra et al. (1997).  With high resolution
imaging we now see that most very red lenses are dominated by 
emission from the lensed images of the host galaxy of the quasar or 
AGN.  For example, the infrared images of MG1131+0456 are dominated
by an extraordinarily bright Einstein ring image of the host whose
properties can be used to prove that the lens galaxy is essentially
transparent with $E(\bv) \ltorder 0.05$ mag (Kochanek et al. 1999). 

Our basic CASTLES survey was not designed to study extinction laws, but
we have begun to obtain detailed IR to UV HST photometry of the lenses 
with significant dust to make the first detailed survey of dust properties
outside the Local Group.  The new data will both greatly expand the
wavelength coverage and minimize the effects of time variability on the
results by obtaining all the data at one epoch.  

\noindent{\bf Acknowledgments:} Support for the CASTLES project was
provided by NASA through grant numbers GO-7495 and GO-7887 from the
Space Telescope Science Institute which is operated by the Association
of Universities for Research in Astronomy, Inc. under NASA contract
NAS 5-26555.  


\end{document}